\documentclass[useAMS,usegraphicx]{mn2e}
\usepackage{rotate}
\usepackage{rotating}
\usepackage{times}
\usepackage{graphicx}
\newif\ifAMStwofonts
\AMStwofontstrue

\def\gs{\mathrel{\hbox{\rlap{\hbox{\lower4pt\hbox{$\sim$}}}\hbox{$>$}}}}
\def\ls{\mathrel{\hbox{\rlap{\hbox{\lower4pt\hbox{$\sim$}}}\hbox{$<$}}}}

\def\einstein{{\it Einstein}}

\def\chandra{{\it Chandra}}
\def\suzaku{{\it Suzaku}}

\def\rosat{{\it ROSAT}}

\def\xmm{{\it XMM-Newton}}

\def\et{{et al.\ }}

\def\mrk335{{Mrk~335}}
\def\1es{{1ES 1927+654}}
\def\rg{{\thinspace r_{\rm g}}}

\def\fvar{{F_{\rm var}}}
\def\chidof{{\chi^2_\nu/{\rm dof}}}
\def\redchi{{\chi^2_\nu}}
\def\delchi{{\Delta\chi^2}}
\def\feka{{Fe~K$\alpha$}}

\def\deg{^{\circ}}

\def\cm{{\rm\thinspace cm}}
\def\erg{{\rm\thinspace erg}}
\def\eV{{\rm\thinspace eV}}

\def\ga{{\rm\thinspace gauss}}

\def\keV{{\rm\thinspace keV}}

\def\s{{\rm\thinspace s}}
\def\ks{{\rm\thinspace ks}}

\def\cts{{\rm\thinspace count}}

\def\cps{\hbox{$\cts\s^{-1}\,$}}

\def\ergpscmps{\hbox{$\erg\cm^{-2}\s^{-1}\,$}}

\def\ergps{\hbox{$\erg\s^{-1}\,$}}

\def\pscm{\hbox{$\cm^{-2}\,$}}

\title[\1es: a bare Seyfert 2]
      {
 \1es: a bare Seyfert 2     }

\author[L. C. Gallo et al.]
       {L. C. Gallo,$^1$
       C. MacMackin,$^1$
       R. Vasudevan,$^2$ 
       E. M. Cackett,$^3$
       A. C. Fabian$^4$
and   F. Panessa$^5$
        \\ 
$^{1}$ Department of Astronomy and Physics, Saint Mary's University, 923 Robie Street, Halifax, NS, B3H 3C3, Canada \\
$^{2}$ Department of Astronomy, University of Maryland, College Park, MD 20742-2421, USA \\
$^{3}$  Department of Physics and Astronomy, Wayne State University, 666 W. Hancock St, Detroit, MI 48201, USA \\
$^{4}$ Institute of Astronomy, University of Cambridge, Madingley Road, Cambridge CB3 0HA\\
$^{5}$ Istituto di Astrofisica Spaziale e Fisica Cosmica (IASF-INAF), via del Fosso del Cavaliere 100, 00133 Roma, Italy \\
}
\date{Accepted. Received. }
\pagerange{\pageref{firstpage}--\pageref{lastpage}}
\pubyear{2012}

\begin{document}
    \maketitle
    \label{firstpage}

    \begin{abstract}
\1es\ is an active galactic nucleus (AGN) that appears to defy the unification model. It exhibits a type-2 optical spectrum, but possesses little X-ray obscuration.  \xmm\ and \suzaku\ observations obtained in 2011 are used to study the X-ray properties of \1es.  The spectral energy distribution derived from simultaneous optical-to-X-ray data obtained with \xmm\ shows the AGN has a typical Eddington ratio ($L/L_{Edd} = 0.014-0.11$).  The X-ray spectrum and rapid variability are consistent with originating from a corona surrounding a standard accretion disc.  Partial covering models can describe the x-ray data; however, the narrow \feka\ emission line predicted from standard photoelectric absorption is not detected.  Ionized partial covering also favours a high-velocity outflow ($v \approx 0.3c$), which requires the kinetic luminosity of the wind to be $\gs 30$ per cent of the bolometric luminosity of the AGN.  Such values are not unusual, but for \1es\ it requires the wind is launched very close to the black hole ($\sim 10\rg$).  Blurred reflection models also work well at describing the spectral and timing properties of \1es\ if the AGN is viewed nearly edge-on, implying that an inner accretion disc must be present.  The high inclination is intriguing as it suggests \1es\ could be orientated like a Seyfert 2, in agreement with its optical classification, but viewed through a tenuous torus.  

    \end{abstract}



    \section{Introduction}
        \label{sect:intro}

        Previous \chandra\ and \rosat\ observations of the active galactic nucleus (AGN)  \1es\ ($z=0.017$) found that it did
        not fit within the standard AGN unification model  (Boller \et 
        2003, hereafter B03).  Its optical spectrum is deficient of the broad emission lines associated with unabsorbed
        Seyfert 1 (Sy1) galaxies, but its X-ray spectrum lacks the absorption associated
        with Seyfert 2 (Sy2) galaxies.  B03 provided a number of possible explanations including an underluminous 
        broad line region (BLR); an X-ray absorber that is optically thick possibly with a higher than usual dust-to-gas
         ratio;  partial covering absorption; or that \1es\ was highly variable, and misclassified due to  non-simultaneous X-ray and optical observations.
        Such ``changing look'' AGN have been previously observed (e.g. Matt \et 2003; Bianchi \et 2005), but simultaneous X-ray and optical observations by Panessa \et (in prep) show this is not the case for \1es\ and that the AGN is a {\it true} optical type 2 (e.g. Bianchi \et 2008, 2012; Panessa \et 2009).  Several objects that exhibited this non-standard behaviour have been discovered (e.g. Panessa \& Bassani 2002;  Mateos \et 2005; Balestra \et 2005; Gallo \et 2006) demonstrating that such objects may not be unusual.  Panessa \& Bassani (2002) speculate that as many as $\sim 10-30$ per cent of optically selected type 2 AGN could exhibit such unorthodox behaviour in X-rays (see also Trouille \et 2009).

	Nicastro (2000) considered the possibility of a class of objects that are truly void of a BLR (i.e. true Seyfert 2s).  
	Assuming the BLR is a wind formed in a region of accretion disc instabilities
        where the disc changes from radiation-pressure dominated  to gas-pressure dominated,
        the radius at which this occurs depends on the accretion rate and will decrease as the accretion
        rate falls.  Once the accretion rate becomes sufficiently low this radius will not be conducive to stable orbits,  the wind will cease, and the BLR will fade.
This is compatible with an evolutionary scenario proposed by Wang \& Zhang (2007), in which they
        postulate that non-HBLR (non-hidden broad line region) Sy2 galaxies, without absorption, are the end state of AGN development.  These galaxies would have  
         the most massive AGNs with the lowest accretion rates.  
It is worth mentioning that the three confirmed true type 2 AGN (NGC~3147, NGC~3660, Q~$2131-427$) all have low Eddington ratios (Bianchi \et 2012).
Tran \et (2011) propose \1es\ is such an object.  
       
       \1es\ may appear atypical even amongst this unusual class.  In the X-rays, Boller (2000) found the object possessed a steep \rosat\ spectrum and large-amplitude variability similar to narrow-line Seyfert 1 galaxies (NLS1s), which are normally associated with high Eddington ratios.   Even Nicastro \et (2003) likened \1es\ to a NLS1.  Wang \et (2012) proposed that \1es\ could be a young AGN that has not yet had time to develop a BLR, and predict that such objects should be distinguished by high Eddington ratios.   
       
       In this work we examine the X-ray properties of \1es\ making use of non-simultaneous \xmm\ and \suzaku\ data obtained in 2011 in order to determine if the X-rays can elucidate the nature of \1es.  While the AGN was discovered in the \einstein\ survey and observed with \rosat\ and the \chandra\ LETG (B03), these most recent observations are the highest quality data obtained to date of \1es\ above $\sim 2\keV$.  In the next section we describe the observations and data processing.   
 In Section~3 we fit the X-ray spectra of \1es\ with traditional models in order to compare with other AGNs, and in Section~4 we model the UV-to-X-ray spectral energy distribution.  We use this information in Section~5 to consider physical motivated spectral models for the X-ray emission.  The X-ray variability over the past 20 years, as well as the shorter time scales during the \suzaku\ and \xmm\ observations, are examined in Section~6.  We discuss our results  and conclusions in Section~7 and 8, respectively.

        \section{Observations and data reduction}
        \label{sect:data}
        \1es\ was observed with \suzaku\ (Mitsuda \et 2007) and \xmm\ (Jansen
        \et 2001) starting on 2011 April 16 and May 20, respectively.  The
        duration of the \suzaku\ observation was $122 \ks$ and that of the \xmm\ observation was $29 \ks$. 
        A summary of the observations is provided in Table~\ref{tab:obslog}.
        
            \begin{table*}
            \begin{center}
            \caption{\suzaku\ and \xmm\ observation logs for \1es.
            The start date of the observation is given in column (1).   The telescope and instrument used is shown in column (2) and (3), respectively. 
            Column (4) is the ID corresponding to the observations.  Good exposure time and source counts (corrected for background) are given in
            column (5) and (6), respectively.  For the EPIC instruments the counts correspond to the $0.3-10\keV$ band.  For the RGS the counts correspond to the $0.3-2\keV$ band.  The values for the \suzaku\ CCDs are taken in the $0.7-1.5$ and $2.5-10\keV$ band.
            }
            \begin{tabular}{cccccc}                
            \hline
            (1) & (2) & (3) & (4) & (5) & (6)  \\
            Start Date   &  Telescope&   Instrument  & Observation ID    &  Exposure & Counts\\
              (year.mm.dd)   &    &    &                    &       (s) &        \\
			\hline
            \hline
            2011.04.16 & \suzaku\  & BI & 706006010  &  71890 & 40611\\
                & &  FI &            &              71900 & 60132 \\
                 & & PIN &          &            74330 & NA \\
            2011.05.20 & \xmm\ & PN    & 0671860201 &  19750 & 99132 \\
                & &  MOS1  &            &             28010 & 35096 \\
                 & &  MOS2  &	         &             28030 & 34728 \\
                 & &  RGS1  &            &             28480 & 4814 \\
                 & & RGS2  &            &             28520 & 5356 \\
            \hline
            \label{tab:obslog}
            \end{tabular}
            \end{center}
            \end{table*}

        The EPIC pn (Str\"uder \et 2001) and MOS (MOS1 and MOS2;
        Turner \et 2001) cameras were operated in small-window and full-window modes,
        respectively, and with the thin filter in place. The Reflection Grating Spectrometers
        (RGS1 and RGS2; den Herder \et 2001) and
        the Optical Monitor
        (OM; Mason \et 2001) also collected data during this time.

        The \xmm\ Observation Data Files (ODFs) from all observations
        were processed to produce calibrated event lists using the \xmm\
        Science Analysis System ({\tt SAS v12.0.0}). 
        EPIC response matrices were generated using the {\tt SAS}
        tasks {\tt ARFGEN} and {\tt RMFGEN}.  Light curves were extracted from these
        event lists to search for periods of high background flaring, which
        was deemed negligible.
        The total amount of good pn exposure is listed in Table~\ref{tab:obslog}.
        Source photons were extracted from a circular region 35$^{\prime\prime}$ across
        and centered on the source.  Pile up was examined for and determined to
        be unimportant.
        The background photons were extracted from an off-source region on the same CCD.
        Single and double events were selected for the pn detector, and
        single-quadruple events were selected for the MOS.
        The MOS and pn data at each epoch are compared for consistency and determined to
        be in agreement within known uncertainties (Guainazzi \et 2010).

        The RGS spectra were extracted using the {\tt SAS} task {\tt RGSPROC} and
        response matrices were generated using {\tt RGSRMFGEN}.
        The OM operated in imaging mode and collected data in the $V$, $UVW1$,
        and $UVM2$ filters.

        During the \suzaku\ observation the two front-illuminated (FI) CCDs
        (XIS0 and XIS3), the back-illuminated (BI) CCD (XIS1), and the HXD-PIN
        all functioned
        normally and collected data.  The target was observed in the
        XIS-nominal position.

        Cleaned event files from version 2 processed data were used in the
        analysis and data
        products were extracted using {\tt xselect}.
        For each XIS chip, source counts were extracted from a $4^{\prime}$
        circular region centred on
        the target.  Background counts were taken from surrounding regions on
        the chip.  Response files
        (rmf and arf) were
        generated using {\tt xisrmfgen} and {\tt xissimarfgen}.      
       After examining for consistency, the data from the XIS-FI were combined
        to create a single spectrum.

        The PIN spectrum was extracted from the HXD data following standard procedures.
        A non-X-ray background (NXB) file corresponding to the observation was
        obtained to generate
        good-time interval (GTI) common to the data and NXB.  The data were
        also corrected for
        detector deadtime.  The resulting PIN exposure was $74330 \s$.  The
        cosmic X-ray background
        (CXB) was modelled using the provided flat response files.  The CXB
        and NXB background files
        were combined to create the PIN background spectrum.  Examination of
        the PIN data yielded no detection of
        the AGN.

        All parameters are reported in the rest frame of the source unless specified
        otherwise.
        The quoted errors on the model parameters correspond to a 90\% confidence
        level for one interesting parameter (i.e. a $\Delta\chi^2$ = 2.7 criterion).
        A value for the Galactic column density toward \1es\ of
        $6.87 \times 10^{20}\pscm$ (Kalberla \et 2005) is adopted in all of the
        spectral fits and solar abundances from Anders \& Grevesse (1989) are assumed unless stated otherwise (see Section~\ref{sect:npc}). 
        K-corrected luminosities are calculated using a
        Hubble constant of $H_0$=$\rm 70\ km\ s^{-1}\ Mpc^{-1}$ and
        a standard flat cosmology with $\Omega_{M}$ = 0.3 and $\Omega_\Lambda$
        = 0.7.  

    \section{Characterizing the X-ray spectra}
    \label{sect:xspectra}
        In this section we attempt to characterize the X-ray spectra of
        \1es\ with traditional phenomenological models.  Initially, the
        spectra from all the \suzaku\ and \xmm\ CCDs were fitted separately.
        It was found that \xmm\ pn data agreed well with the MOS data.
        Similarly, it was found that the data from the \suzaku\ FI detectors
        agreed well with data from the BI CCD.
        Since all the data agreed within the calibration uncertainties we
        present only the \xmm\ pn and \suzaku\  FI spectra for multi-epoch comparison and for 
        ease of presentation.   Due to uncertainties in the calibration the FI
        data are ignored below $0.7\keV$ and between $1.5-2.5\keV$.  
        In general, given the AGN spectrum is rather steep the data quickly become noisy at higher energies.
        This seems to be the main cause of the increased residuals above $\sim 7\keV$.
        The \suzaku\ data
        above $9\keV$ become completely background dominated and are consequently
        ignored.  The pn data are used between $0.3-10\keV$.

\begin{figure}
            \rotatebox{0}
            {\scalebox{0.55}{\includegraphics{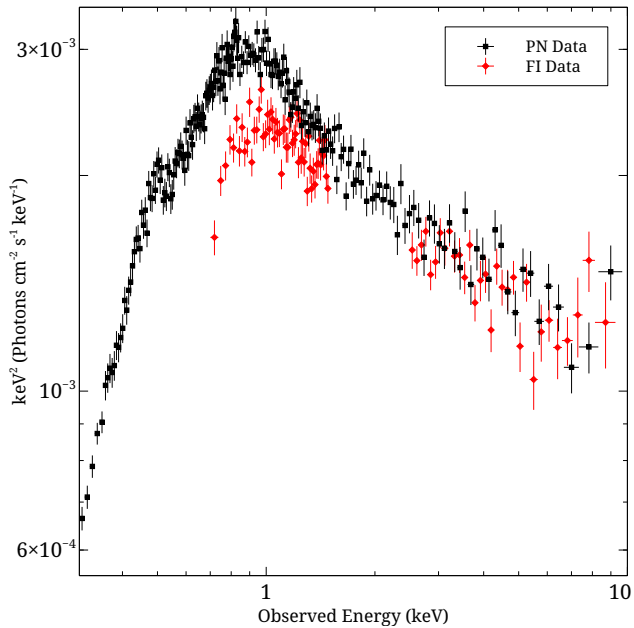}}}
            \caption{The \xmm\ pn (black) and \suzaku\ FI (red) spectra corrected for instrumental differences.
            Both spectra exhibit a steep slope and a drop at low energies, likely from absorption in addition to Galactic levels.
            }
            \label{fig:eeuf}
            \end{figure}

        For comparison the pn and FI spectra are shown in
        Fig~\ref{fig:eeuf} corrected for instrumental differences. The
        spectra are comparable at both epochs except that the \suzaku\ spectrum appears slightly flatter and dimmer. 
        Both spectra show a drop at lower energies commensurate with
        some level of absorption in addition to Galactic.
        
        \subsection{The $2.5-10\keV$ band}
        \label{subsect:highe}
            Fitting the $2.5-10\keV$ band at each epoch with a single power law
            modified by the Galactic column density resulted in a good fit
            ($\chi^{2}_{\nu}=0.99$).  The photon indices were $\Gamma \approx 2.41$ and
            $\Gamma \approx 2.27$ for the pn and FI, respectively.   If the X-ray
            spectra were highly absorbed like in typical Seyfert 2s one would predict a strong narrow \feka\
            feature at around $6.4\keV$, however the residuals show no deviations
            around this band.  Adding a narrow ($\sigma=1\eV$) Gaussian profile at
            $6.4\keV$ to the model did not provide a significant improvement.  The
            upper-limit on the flux of the narrow feature is 
            $< 9.39 \times 10^{-15} \ergpscmps$, corresponding to an upper-limit on the equivalent width of $EW < 30 \eV$
            and $EW < 28 \eV$ for pn and FI, respectively. 
            Allowing the width and energy of the Gaussian profile to vary freely
            did not improve the fit over the simple power law.

        \subsection{The broadband X-ray spectra}
        \label{subsect:bbs}

\begin{figure}
                \rotatebox{0}
                {\scalebox{0.55}{\includegraphics{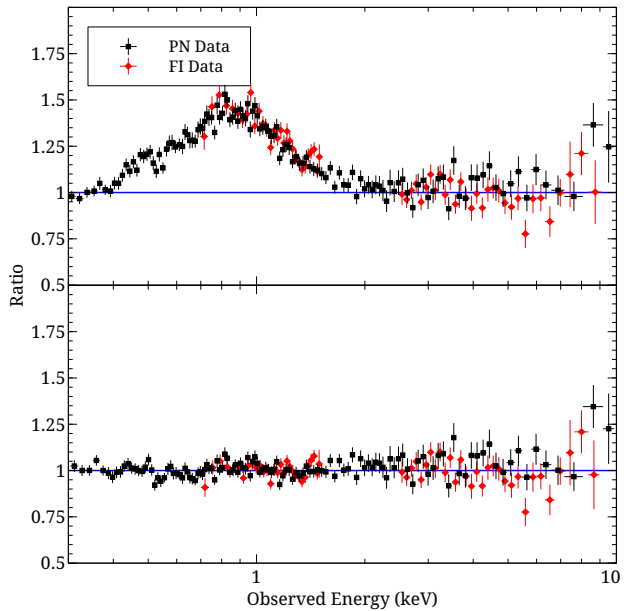}}}
                \caption{Top panel: The residuals remaining when fitting a power law absorbed by Galactic column density to the $2.5-10\keV$ band and extrapolating to $0.3\keV$.  A soft-excess below $\sim 2 \keV$ is revealed, which is then modified at energies below $\sim0.7\keV$.  Lower panel: The residuals after the addition of blackbody modified by additional absorption. 
                  }
                \label{fig:softexc}
                \end{figure}

            When the $2.5-10\keV$ power law fit from Section~\ref{subsect:highe} is
            extrapolated to lower energies, a clear soft excess is evident below
            $\sim2\keV$ (Fig~\ref{fig:softexc} top panel).

            Adding a blackbody component worked well to describe the
            data. 
            Since the pn data extend to lower energies than the FI data, the blackbody temperature was linked between the two
            epochs, but all other parameters were allowed
            to vary.  The fit was statistically acceptable with  $\chidof = 1.01/1226$.  
            Replacing the blackbody with a second power
            law (i.e. a double power law) or refitting the spectra with a broken power law  generated poorer fits 
            ($\chidof =  2.20/1226$ and $1.20/1227$, for the double power law and broken power law, respectively). 
                  
            Although reasonable fits were possible, in particular with the blackbody plus power law model, all three models showed  
            a drop in the residuals at energies below about $0.5\keV$.  The addition of a neutral absorber intrinsic to the host galaxy
            ({\tt ztbabs}) improved the residuals in all cases.  The level of absorption was modest ($\sim 5-10 \times
            10^{20} \pscm$) and it was in line with what was reported with \rosat\ observations (B03), which were sensitive to even lower energies than the pn.  The parameters are reported for each model  in Table~\ref{tab:traditionalModels}.  The residuals from the best-fitting blackbody plus power law model are shown in Fig~\ref{fig:softexc} (lower panel).  The observed $0.3-2\keV$ and $2-10\keV$ fluxes during the \xmm\ observation are 6.8 and $3.7 \times 10^{-12} \ergpscmps$, respectively.  The intrinsic $2-10\keV$ luminosity, corrected for absorption in the Galaxy and host galaxy, is $L  = 2.4 \times 10^{42} \ergps$ during the \xmm\ observation.

                \begin{table*}
                \begin{center}
                \caption{Results from fitting \xmm\ and \suzaku\ spectra with traditional models.   The broadband X-ray model is stated in column (1). 
               The model components and parameters are shown in column (2) and (3), respectively.  The parameter values during the \xmm\ and \suzaku\ epochs are reported in columns (4) and (5), respectively.                
                Observed fluxes are reported for each component in units of $\times 10^{-12} \ergpscmps$ over the $0.3-10\keV$ band except in the case of the broken power law model where fluxes are reported above and below the break energy.  Parameters that are linked between epochs are only reported in one column.  The Galactic column density has been included in all models.
                }
            \begin{tabular}{ccccc}                
\hline
(1) & (2) & (3) & (4) & (5) \\
Model & Model component & Model parameter  & \xmm & \suzaku \\
\hline                
Broken power law & Intrinsic Absorption & $N_{H}$  ($\pscm$) & $1.24\pm0.08 \times 10^{21}$ & \\
  & Power law & $\Gamma_1$ & $3.30\pm0.06$ & \\
   & & Flux  & $25.3\pm0.1$ & $21.3\pm0.3$ \\
   & & $E_{b}$ (\keV) & $1.82\pm0.07$ & \\
   & & $\Gamma_2$ & $2.47\pm0.04$ & $2.27\pm0.04$ \\
   & & Flux  & $4.06\pm0.02$ & $3.99\pm0.03$ \\
& Fit Quality & $\chidof$ &  $1.11/1226$ &  \\
\hline
Double power law & Intrinsic Absorption & $N_{H}$ ($\pscm $) & $1.22^{+0.11}_{-0.10} \times 10^{21}$ & \\
& Power law 1 & $\Gamma$ & $3.40^{+0.12}_{-0.11}$ & \\
    & & Flux  & $26.2\pm0.2$ & $20.6\pm0.2$ \\
& Power law 2 & $\Gamma$ & $1.49^{+0.17}_{-0.20}$ & $1.60^{+0.11}_{-0.13}$ \\
    & & Flux & $3.13\pm0.10$ & $4.06\pm0.10$ \\
& Fit Quality& $\chidof$ &  $1.23/1225$  & \\
\hline
Blackbody plus  & Intrinsic Absorption & $N_{H}$ ($\pscm$) & $2.58\pm0.01 \times 10^{20}$ & \\
power law & Power law & $\Gamma$ & $2.39^{\pm}0.04$ & $2.27^{\pm}0.04$ \\
    & & Flux  & $3.35\pm0.06$ & $2.75\pm0.10$ \\
& Blackbody & $kT$ (\keV) & $0.170{\pm}0.005$ &  \\
    & & Flux  & $12.36\pm0.09$ & $10.7\pm0.1$ \\
& Fit Quality & $\chidof$ & $0.97/1225$ & \\
\hline
\label{tab:traditionalModels}
\end{tabular}
                \end{center}
                \end{table*}


        \subsection{RGS data}
        \label{subsect:RGS}
            The small level of cold absorption evident in the CCD
            spectra motivated investigation at higher spectral resolution with the
            \xmm\ RGS data.  The RGS data between 0.4--2.0\keV\ were examined 
            100 channel at a time and fitted with a simple power law (corrected for
            Galactic absorption).   A Gaussian profile was used to examine for
            possible narrow emission and absorption features.    A potential
            absorption feature was found around $0.85\keV$, but was not
            constrained in the error analysis.  No other features were detected
            within the available signal-to-noise.
           
         \section{Optical-to-X-ray spectral energy distribution}
        \label{subsect:SED}
        We construct a broadband SED using the \xmm\ pn and OM 
data along with the \suzaku\ FI data.  
The OM points are corrected for Galactic dust extinction
[$E(B-V)_{\rm Gal} = 0.087 \pm 0.003$] (Schlegel, Finkbeiner \& Davis 1998)\footnote{http://irsa.ipac.caltech.edu/applications/DUST/} before fitting in \textsc{xspec}.  
We follow the
procedure of Vasudevan et al. (2009) for constructing the SED, using
the {\tt xspec} model combination \textsc{zdust(diskpn)+ztbabs(bknpower)},
to allow for
optical/UV reddening as well as X-ray absorption obscuring the intrinsic accretion disc and power-law emission, respectively.
 The power law
falls in the UV where the disc dominates.  We include an intrinsic
(host galaxy) dust reddening component of $E(B-V) = 0.55$, corresponding
to the upper limiting intrinsic extinction $A_{V}$ of 1.71 identified
in B03 from the $H_{\alpha}/H_{\beta}$ line ratio,
using the standard relation $E(B-V) = A_{V} / R_{V}$ and assuming
$R_{V}=3.1$.   We freeze the normalisation of the \textsc{diskpn}
model using the black hole mass estimate of log($M_{\rm BH}/M_{\rm
sun}$)=7.34 from Tran et al. (2011), and assume an accretion disc
extending close to the innermost stable orbit, down to 6 $R_{\rm g}$.
The photon index of the power law in the 2--10 keV regime is fixed at
2.3, as found from the analysis in Section~\ref{subsect:bbs}  Under these assumptions, we
obtain a bolometric (absorption-corrected) luminosity of $3.0 \times
10^{44} \rm erg \thinspace s^{-1}$ (integrated between 0.001 and 100
keV), and find that at least 53 per cent of this power is absorbed by
dust.  These assumptions yield an Eddington ratio of $0.11$, larger
than the value of 0.006 estimated from Tran et al. (2011) due to the extra disc
contribution extrapolated here into the unobservable far-UV.   If we
consider an alternate disc geometry where the disc is truncated
further out from the black hole (at, e.g. 60 $R_{\rm g}$), we find
that the Eddington ratio reduces to 0.046 (with 80 per cent of the
bolometric flux absorbed by dust reddening), but is still well above
the Tran et al. estimate.  If we simply extrapolate a power law from
the X-rays down to the optical and UV, discarding any contribution
from the disc and integrate over the same range, we find a lower limit
to the Eddington ratio of 0.014.  It is also possible that the
optical/UV flux is not from a canonical accretion disc, but we do not
explore that possibility further here.

Overall, this implies that the accretion rate of the object ($0.014-0.11$) is in
between the Wang et al. and Tran et al. estimates,
regardless of the type of geometry we assume for the accretion disc.
A substantial fraction of the intrinsic luminosity is probably
absorbed by dust (between 50-80 per cent).  We have not attempted a
host-galaxy light correction to the XMM-OM fluxes, but this would serve to
push the true bolometric luminosity down and reduce the Eddington
accretion rate from the values seen here.  The optical images reveal a
crowded field, so a more refined analysis may be needed to more
accurately estimate the bolometric luminosity and Eddington ratio.

             \begin{figure}
             \rotatebox{0}
             {\scalebox{0.55}{\includegraphics{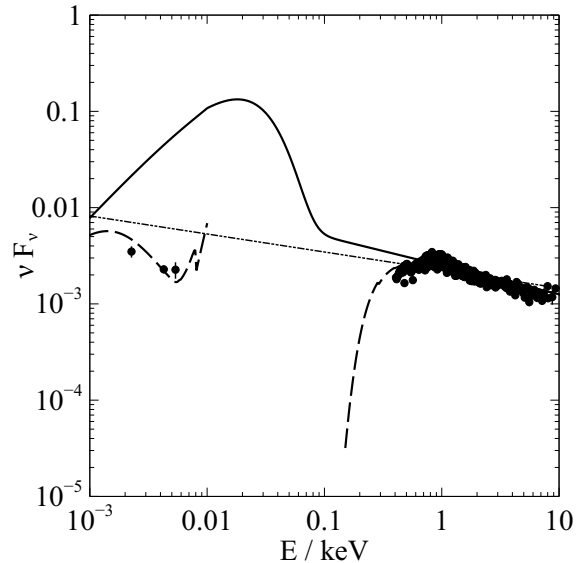}}}
             \caption{Spectral energy distribution (SED) for 1ES 1927+654, from \xmm\ PN and
OM data and \suzaku\ FI data (black filled points).  The dashed line
shows an accretion disk and broken power law model fit with dust
extinction in the optical--UV and photoelectric absorption in X-rays,
and the solid line shows the unabsorbed fit (with dust reddening and
absorption set to zero), with a prominent unobservable accretion disc
component.   The dot-dashed line shows a simple power law
extrapolation from X-rays down to optical wavelengths.
             }
             \label{fig:lumabs}
             \end{figure}

	\section{Multi-epoch spectral analysis of 1ES 1927+654}
	\label{sect:physicalModels}

	Based on our phenomenological approach in Section~\ref{sect:xspectra},
	there are only slight differences between the \xmm\ and \suzaku\ data sets (Table~\ref{tab:traditionalModels}).  
	In this section we attempt to model the 
	two broadband X-ray data sets simultaneously and in a self-consistent manner while testing more physically motivated models.  
	For example, while the blackbody plus power law model presented in Section~\ref{sect:xspectra} was statistically pleasing, the implied thermal origin for
	the soft excess is questionable.  Given the relatively large black hole mass, even for the highest  Eddington accretion rate measured in Section~\ref{subsect:SED}, a standard accretion disc extending down to $2\rg$ would peak at a temperature of $kT\approx45\eV$.  Any blackbody disc component
	in \1es\ should have a temperature that is less than one-quarter of what we measure in the X-rays.   
	In general, the lack of T$\propto {\rm M}^{-1/4}$ relation between the black hole
	mass and disc temperature observed in samples of unobscured AGN (e.g. Crummy \et 2006) suggests a non-thermal
	origin for the soft excess.  Both absorption (e.g. Gierli\'nski \& Done 2004) and blurred reflection models (e.g. Crummy \et 2006) can be used to describe
	the shape of the soft excess via atomic processes, and we will attempt such models to describe the spectrum of \1es. 
	In doing so, we will also be testing some of the scenarios that have been put forth to describe the 
	characteristics of \1es.

	\subsection{Neutral partial covering}
	\label{sect:npc}
	
    	We consider the scenario in which the intrinsic power law continuum is partially obscured by a neutral 
    	absorber that lies along the line-of-sight.  The observed X-ray emission is then the combination of the 
    	primary emitter and a highly obscured component.  Such models have been proposed for type-1 Seyfert 
    	galaxies with relative success in fitting the spectrum (e.g.  Gallo \et 2004; Tanaka \et 2005; Grupe \et 2008).
	
\begin{figure}
			\rotatebox{0}
			{\scalebox{0.32}{\includegraphics{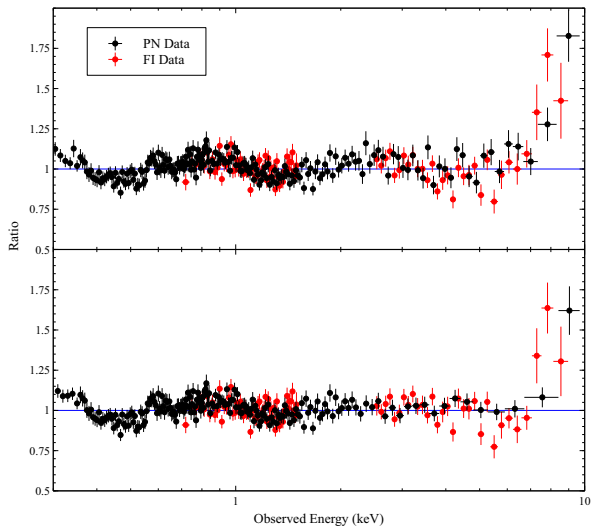}}}
			\caption{Top panel: The residuals remaining from fitting the \xmm\ and \suzaku\ spectra with one neutral absorber partially covering the power law source.  The changes are attributed to variations in the covering fraction and column density at each epoch.  
			Lower panel: The residuals remaining from fitting the \xmm\ and \suzaku\ spectra with two neutral absorbers partially covering the power law source. In this case the variations from one epoch to the next are attributed to changes in the covering fraction of only one absorber.
See text and Table~\ref{tab:modelData} for details.
			}
			\label{fig:npc}
			\end{figure}

		Since the \xmm\ and \suzaku\ observations were separated by only about one month, we assumed that the 
		primary emitter (e.g. the power law component) remained constant in shape and normalisation and that only changes in the absorber 
		(i.e. the covering fraction and column density) would be necessary to describe the differences at the two epochs.  
		The simplest case of the single absorber resulted in a mediocre fit ($\chidof = 1.18/1226$; Fig~\ref{fig:npc} top panel).  
		The addition of a second absorber (i.e. neutral double partial covering) was favorable, as an improvement 
		could be achieve by changing only the covering fraction of the second absorber ($\delchi = 43$ for 1 
		additional free parameter).  The model and quality of fit are shown in 
		Table~\ref{tab:modelData} and Fig~\ref{fig:npc} (lower panel), respectively.

			\begin{figure}
			\rotatebox{0}
			{\scalebox{0.32}{\includegraphics{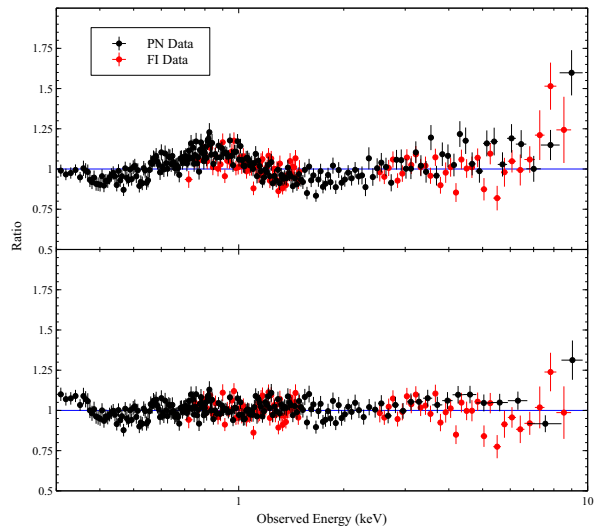}}}
			\caption{Top panel: The residuals remaining from fitting the \xmm\ and \suzaku\ spectra with one ionised absorber partially covering the power law source.   The ionisation parameter, covering fraction and column density of the absorber are allowed to vary at each epoch.
			Lower panel: The improved residuals from the addition of a second ionised absorber are shown.  The variations from one epoch to the next are attributed to changes in the covering fraction of each ionised absorber.  See text and Table~\ref{tab:modelData} for details.
			}
			\label{fig:ipc}
			\end{figure}

	The two absorbers have considerably different densities, which could be indicative of a density gradient along
	the line-of-sight as opposed to two distinct absorbing regions (e.g. Tanaka \et 2004).  
The slightly lower flux during the \suzaku\ observation can be attributed to slightly higher covering fraction while the 
	primary 
	power law emission remains unchanged between the epochs.  Allowing the power law component to vary along with the 
	absorbers does not produce a significant improvement.
	
	While the fit is reasonable, there are clearly spectral regions where the model does not adequately fit the data, 
	for example between $0.4-0.6\keV$ and $7-10\keV$ (Fig~\ref{fig:npc} top panel). 
	The $0.4-0.6\keV$ residuals can be improved with the addition of an absorption edge with rest-frame energy $E=0.39\pm0.01\keV$ and optical depth $\tau=0.66\pm0.12$.  Given the energy the feature seems unlikely to originate from calibration issues around the oxygen edge.
	We considered the possibility that the residuals, could be improved by adjusting the elemental abundance.
	In other abundance tables, like those of Wilms, Allen \& McCray (2000) or Grevesse \& Sauval (1998), the oxygen abundance relative to hydrogen
	can be almost half of what is used in Anders \& Grevesse (1989).  However, while the adopted abundance table does influence some model parameters, the residuals between $0.4-0.6\keV$ do not change significantly.

	\subsection{Ionized partial covering}
	\label{sect:ipc}
	
	The negative residuals between $0.4-0.6\keV$ resulting from the neutral partial covering model (Fig~\ref{fig:npc}) 
	could be indicative of an absorber with some level of ionisation.  A moderately ionised
	absorber will preferentially remove intermediate energy X-rays giving rise to a spectral break in the $0.3-10\keV$ band generating a soft-excess
	and it could account for some of the low-energy residuals. 
	
	As with single neutral partial covering, a single ionised absorber ({\tt zxipcf} in {\tt xspec}) was a modest fit to the data 
	($\chidof = 1.32/1225$; Fig~\ref{fig:ipc} top panel).  A considerably better fit could be be achieved with the addition of a second 
	ionised absorber.  We tested various combinations of parameters to determine which were the most important to 
	describe the spectra in a self-consistent manner.  We determined that an excellent fit could be obtained 
	($\chidof = 1.01/1221$) when the covering fraction of each absorber was allowed to vary at each epoch
	(Table~\ref{tab:modelData} and Fig~\ref{fig:ipc} lower panel), but we also recognize the power law photon index is considerably steeper than is normally seen in AGN (likewise for the neutral partial covering model).
	In addition, we found the best fit was achieved when the absorbers were significantly blueshifted with respect to the source 
	frame, consistent with an outflow of $90 000 \pm 3000$ km/s ($0.30 \pm 0.01$ c). 
	The value of the blueshift was tightly constrained independent of the uncertainty in other covering parameters like the column density,
	covering fraction, and ionisation.

         We note the ionisation parameter of the second absorber is rather low log ($\xi$) $\approx -0.54$ and could perhaps be described with another
         neutral absorber instead (e.g. {\tt ztbabs}).  The fit was significantly poorer than the double ionised absorber, but acceptable ($\chidof=1.05/1222$). 
         We do not test various combinations of cold and warm absorbers any further.  In all likelihood, the absorbers may not be distinct regions with different densities,  but a single region with some density and/or ionisation gradient.
	Comparing the residuals from the ionized and neutral partial covering models (Fig~\ref{fig:npc} and \ref{fig:ipc}) shows the ionized partial covering is favoured.  The deviations between $0.4-0.6\keV$ and $7-10\keV$ seen in the neutral absorber model are considerably improved with the ionized absorber.

If the partial covering model were correct it would only have a small impact on the Eddington ratio estimated in Section~\ref{subsect:SED}.
The unabsorbed X-ray luminosity in this model is approximately $3.5$ times greater than the value used in Section~\ref{subsect:SED}, thereby increasing the bolometric luminosity only slightly.  The unabsorbed putative disc would still dominate and the upper limit on the Eddington ratio would only increase to about $0.13$.

	\subsection{Compton-thick absorption}
	\label{sect:ct}
\begin{figure}
			\rotatebox{0}
			{\scalebox{0.55}{\includegraphics{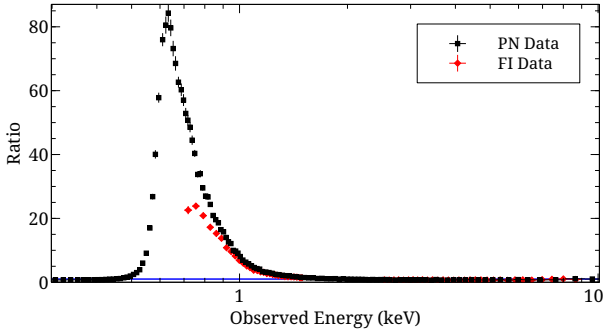}}}
			\caption{The residuals based on an edge-on dusty torus model fitted to the data.
			}
			\label{fig:mytorus}
			\end{figure}

	There has been speculation that  \1es\ could be a Compton-thick source.  Indeed, Bianchi \et (2012) found that some true Sy~2 candidates were in fact Compton-thick.  With observations available in the $0.3-50\keV$ 
	band (i.e. including the PIN null detection) we can test this model more robustly.  For this purpose we used the model {\tt MYTorus} (Murphy \& Yaqoob 2009), 
	in which we view the central engine through some fraction of the torus.  Reasonable fits were established, but only in 
	cases with low, face-on, inclinations ($i \approx 0\deg$), which would be indicative of no absorption from a torus.   Attempts to fix the disc inclination more
	edge-on ($i\approx 85\deg$) generated very poor fits (Fig~\ref{fig:mytorus}).
	
	\subsection{Ionized Disc Reflection}
	\label{sect:bbc}
        Reflection from an ionized disc blurred for relativistic effects close to the black hole is often adopted 
        to describe the origin of the soft excess (Ballantyne, Ross \& Fabian 2001; Ross \& Fabian 2005), and has 
        been successfully fitted to the spectra of unabsorbed AGN (e.g. Fabian et al. 2004; Crummy et al. 2006; Ponti 
        et al. 2010; Walton et al. 2013).  We consider this scenario to describe the spectra of \1es\ and the 
        variations between the two epochs.
        
        Given that the observations were obtained within about one-month of each other and that the AGN is in a similar 
        flux state we do not expect significant variations.  Initially, only the power law slope and normalisation; 
        and the reflector ionisation and normalisation were permitted to vary.  The blurring parameters were fixed to 
        default values and linked between the two observations.  The model produced a reasonable fit ($\chidof=1.04/1224$).  
        
\begin{figure}
			\rotatebox{0}
			{\scalebox{0.55}{\includegraphics{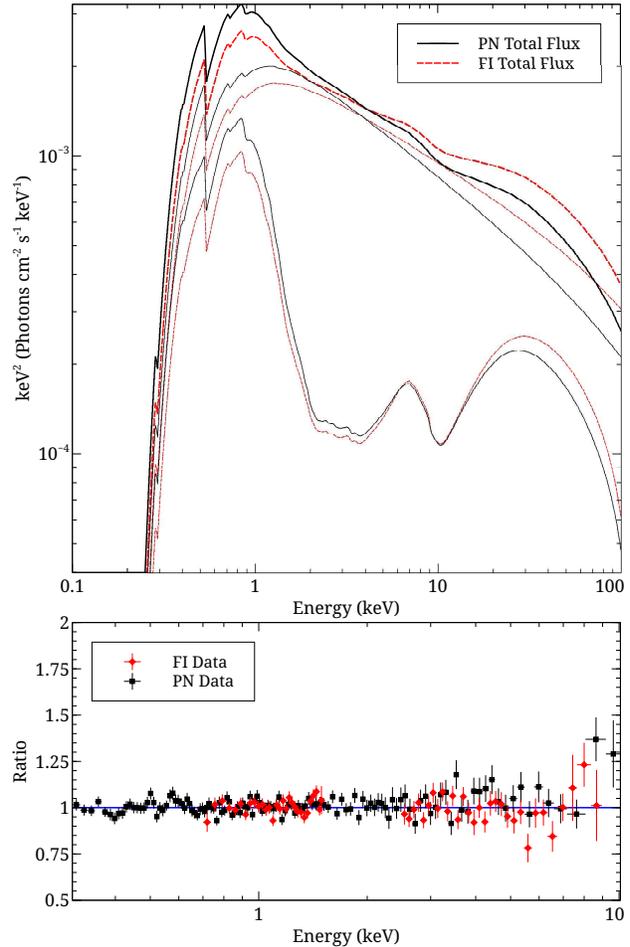}}}
			\caption{Top panel: The blurred reflection model (A) for the \suzaku\ FI (red) and \xmm\ (black) data.
			The source is power law dominated over the entire X-ray band at both epochs.  Lower panel: The residuals from the model fit in the $0.3-10\keV$ band.
			}
			\label{fig:reflmo}
			\end{figure}

\begin{figure}
			\rotatebox{270}
			{\scalebox{0.35}{\includegraphics{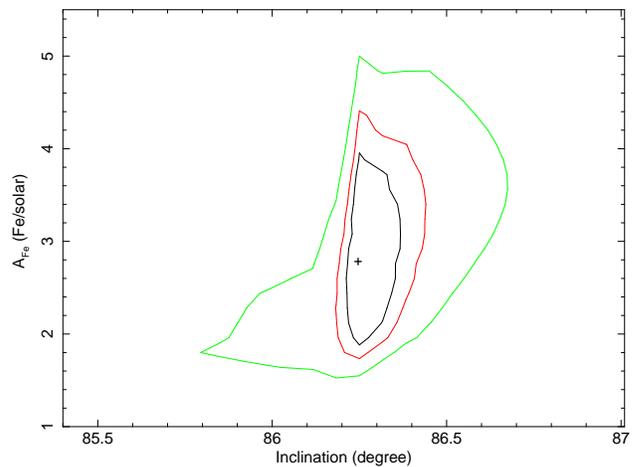}}}
			\caption{A contour plot of the blurred reflection model (A) fit with varying inclination and iron abundance. The contour
			lines are drawn around $\Delta \chi^2$ values of 3, 5, and 9.
			}
			\label{fig:cont}
			\end{figure}

        We examined each individual parameter and then combinations of parameters, to determine which variables would improve the fit most significantly.    We determined that the fit was substantially improved 
        ($\delchi=104$ for 2 additional free parameters) only when the inclination and iron abundance were free 
        to vary (Fig~\ref{fig:reflmo} and see Blurred Reflection (A) in Table~\ref{tab:modelData}).  The iron abundance was found 
        to be about twice the solar 
        value ($A_{Fe}=2.78^{+1.03}_{-1.21}$) and the disc was significantly inclined (edge-on, $i=86\pm1\deg$).  Permitting more than these two additional parameters to vary did not improve the quality of the fit.
                We further 
        examined how the two parameters depend on each other and found that they influence each other rather modestly 
        (Fig~\ref{fig:cont}).

        We note that the measured inclination is rather extreme while the emissivity index is fixed to $q=3$ as expected from lamp-post illumination at a large distance.  However, the quality of the fit does not change if the 
         emissivity is fixed to values of $q>3$.  Similarly, while the disc inner edge was initially fixed to $R_{in}=4.5\rg$,  the value in the current fit could be reduced to $R_{in}=3.3\rg$ and still achieve a perfectly acceptable fit ($\redchi=0.99$).  
         
         In fact, a second blurred reflection model that is more relativistic in nature fits the spectra equally well.  In this case, we fixed the inner emissivity 
         index to $q_{in}=5$ and the disc inner edge  to $R_{in}=1.5\rg$ (see Blurred Reflection (B) in Table~\ref{tab:modelData}).  While the iron abundance and ionization parameter slightly decrease in this relativistic fit, the inclination still remains high ($i\approx 90\deg$).  
         
         The two blurred reflection models demonstrate that compact geometries for the primary emitter cannot be ruled out, but the fit shows a preference for high inclinations.  We also note there are no dominant \feka\ emission features in the spectrum, so the blurred reflection models are primarily fitting the soft excess.

                \begin{table*}
                \begin{center}
                \caption{Results from fitting the \xmm\ and \suzaku\ spectra with physically motivated models.   The model is stated in column (1). 
               The model components and parameters are shown in column (2) and (3), respectively.  The parameter values during the \xmm\ and \suzaku\ epochs are reported in columns (4) and (5), respectively.                
                  Parameters that are linked between epochs are only reported in one column.  The Galactic column density has been included in all models.  
                }
                \begin{tabular}{ccccc}                
                \hline
                (1) & (2) & (3) & (4) & (5) \\
                Model & Component &  Parameter  & \xmm & \suzaku \\
                \hline
                Neutral double partial covering &  Intrinsic Absorption & $N_{H}$ (\pscm) & $(1.21 \pm 0.08) \times 10^{21}$ & \\
	    		 & Absorber 1 & $N_{H}$ (\pscm) & $5.8^{+3.3}_{-2.0} \times 10^{23}$ & \\
	    		 & & $C_f$ & $0.52^{+0.11}_{-0.09}$ & \\
	    		 & Absorber 2 & $N_{H}$ (\pscm) & $(7 \pm 1) \times 10^{22}$ & \\
	    		 & & $C_f$ & $0.57^{+0.03}_{-0.04}$ & $0.64 \pm 0.03$ \\
	    		 & Power Law & $\Gamma$ & $3.27 \pm 0.06$ &  \\
	    		 & Fit Quality & $\chidof$ & $1.15/1225$ & \\
                \hline
                Ionized double partial covering & Intrinsic Absorption & $N_{H}$ (\pscm) & $9.23^{+0.40}_{-1.16} \times 10^{20}$ & \\
			     & Absorber 1 & $N_{H}$ (\pscm) & $6.6^{+1.6}_{-0.7} \times 10^{23}$ & \\
			     & & $\log(\xi)$ & $3.03^{+0.11}_{-0.16}$ & \\
			     & & $C_f$ & $0.49 \pm 0.11$ & $0.34 \pm 0.09$ \\
			     & & $z$$^1$ & $-0.30 \pm 0.01$ & \\
			     & Absorber 2 & $N_{H}$ (\pscm) & $3.34^{+2.5}_{-1.0} \times 10^{22}$ & \\
			     & & $\log(\xi)$ & $-0.54^{+0.81}_{-0.70}$ & \\
			     & & $C_f$ & $0.28^{+0.11}_{-0.08}$ & $0.43^{+0.07}_{-0.11}$ \\
			     & & $z$$^1$ & $-0.30 \pm 0.01$ & \\
			     & Power Law & $\Gamma$ & $2.78 \pm 0.07$ &  \\
			     & Fit Quality & $\chidof$ & $1.01/1221$ & \\
			    \hline              
			    Blurred reflection (A) & Intrinsic Absorption & $N_{H}$ (\pscm) & $5.00^{+0.39}_{-0.56} \times 10^{20}$ & \\
			     & Power Law & $\Gamma$ & $2.48^{+0.05}_{-0.03}$ & $2.36 \pm 0.04$ \\
			     & & $E_{cut}$ & $300^f$ & \\
			     & Blurring & $q$ & $3^f$ & \\
			     & & $R_{in}$ ($r_g$) & $4.5^f$ \\
			     & & $R_{out}$ ($r_g$) & $400^f$ & \\
			     & & $i$ ($\deg$) & $86 \pm 1$ & \\
			     & Reflection & $A_{Fe}$ ($Fe/solar$) & $2.78^{+1.03}_{-1.21}$ & \\
			     & & $\xi$ & $384^{+162}_{-131}$ & $310^{+197}_{-91}$ \\
			     & Fit Quality & $\chidof$ & $0.96/1222$ & \\
			    \hline
			    Blurred reflection (B) & Intrinsic Absorption & $N_{H}$ (\pscm) & ($4.40\pm 0.05) \times 10^{20}$ & \\
			     & Power Law & $\Gamma$ & $2.43 \pm 0.05$ & $2.31 \pm 0.05$ \\
			     & & $E_{cut}$ & $300^f$ & \\
			     & Blurring & $q_{in}$ & $5^f$ & \\
			     & & $R_{in}$ ($r_g$) & $1.5^f$ \\
			     & & $R_{out}$ ($r_g$) & $400^f$ & \\
			     & & $R_{break}$ ($r_g$) & $5^f$ & \\
			     & & $q_{out}$ & $3^f$ & \\
			     & & $i$ ($\deg$) & $90^{+0}_{-1}$ & \\
			     & Reflection & $A_{Fe}$ ($Fe/solar$) & $0.76^{+0.37}_{-0.16}$ & \\
			     & & $\xi$ & $200^{+25}_{-88}$ & $202^{+35}_{-120}$ \\
			     & Fit Quality & $\chidof$ & $0.96/1222$ & \\
		\hline
                \label{tab:modelData}
                \end{tabular} \\
                $^1$The redshifts of the partial covering components in this model are linked.
                \end{center}
                \end{table*}


    \section{Timing analysis}
    \label{sect:light curves}

\begin{figure}
                \rotatebox{0}
                {\scalebox{0.55}{\includegraphics{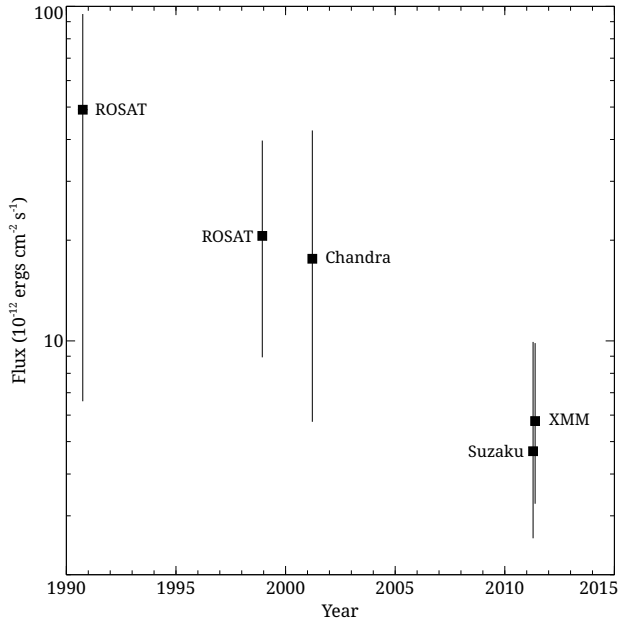}}}
                \caption{The $0.5-2\keV$ light curve of \1es\  over the past $\sim 20$ years.  The mission from which the data are obtain is labeled beside each point.  The vertical bars represent the range between minimum and maximum
flux that is observed at that epoch.
                }
                \label{fig:LongCurve}
                \end{figure}

           In Fig~\ref{fig:LongCurve} the  $0.5-2\keV$ light curve of \1es\ over the past $\sim 20$ years is shown.  The vertical bars mark the range between minimum and maximum flux that is observed at each epoch.
           The \xmm\ and \suzaku\ fluxes are measured from the blackbody plus power law model described in Section~\ref{subsect:bbs}.  Fluxes from the earlier data are estimated from the count rates and models presented by B03 using {\tt WebPIMMS}.\footnote{http://heasarc.gsfc.nasa.gov/Tools/w3pimms.html}

\1es\ varies significantly during all observations.  
On average, the \xmm\ and \suzaku\  observations catch \1es\ at lower X-ray flux levels than have
been previously observed.

        In this section we examine the variability of the AGN during the \xmm\ and \suzaku\ observations.
        The  short, uninterrupted \xmm\ observations will provide the opportunity to examine the rapid 
        variability (i.e. over hundreds of seconds) with high count rates.  The \suzaku\ observation 
        is interrupted every $\sim5.7\ks$ due to Earth occultation, but the longer duration of the observation 
        allows us to study the variability over $\sim1.5$-days.  

        \subsection{Rapid and large amplitude variability}
            One of the distinguishing characteristics of \1es\ from \rosat\ observations was the rapid variability, comparable 
            to what is seen in narrow-line Seyfert 1 galaxies (Boller 2000; B03).   Even at this 
            lower flux level, the AGN continues to exhibit impressive variability. 

\begin{figure}
            \rotatebox{0}
            {\scalebox{0.55}{\includegraphics{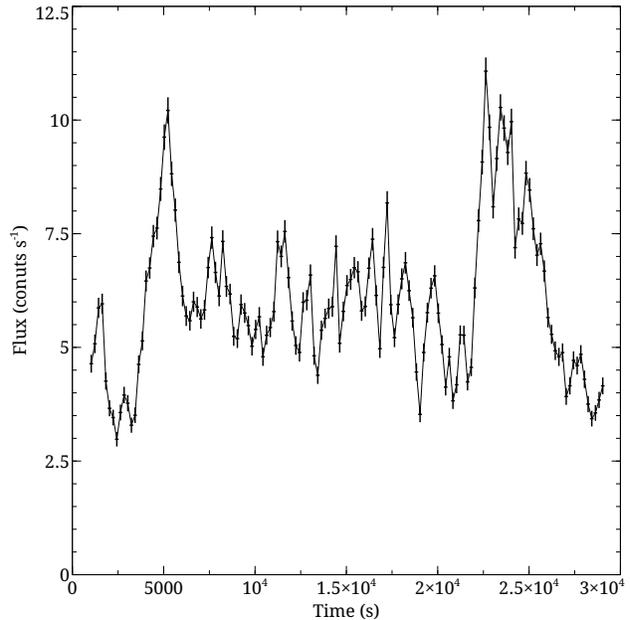}}}
            \caption{The $0.2-12\keV$ \xmm\ light curve binned in 200\s\ intervals. 
            }
            \label{fig:XMMCurve}
            \end{figure}

            The broadband ($0.2-12\keV$) pn light curve is shown in Fig~\ref{fig:XMMCurve}.  The 
            variability is persistent over the $30\ks$.  During the largest flares the count rate 
            changes by more than a factor of 3 in about a thousand seconds.  We calculated the radiative efficiency 
            ($\eta$) assuming photon diffusion through a spherical mass of accreting material (Fabian 1979).  
            The most rapid rise seen in the light curve corresponds to a luminosity change of $8.95 \times 10^{42} \ergps$ in about  
            900 s (rest frame).  However, as \1es\ is a relatively low-luminosity AGN the lower limit on the radiative 
            efficiency is $\eta \ge 0.5$ per cent.  Anisotropic emission or a maximum spinning black hole is not required to describe the 
            large amplitude variability in \1es. 

\begin{figure}
            \rotatebox{0}
            {\scalebox{0.48}{\includegraphics{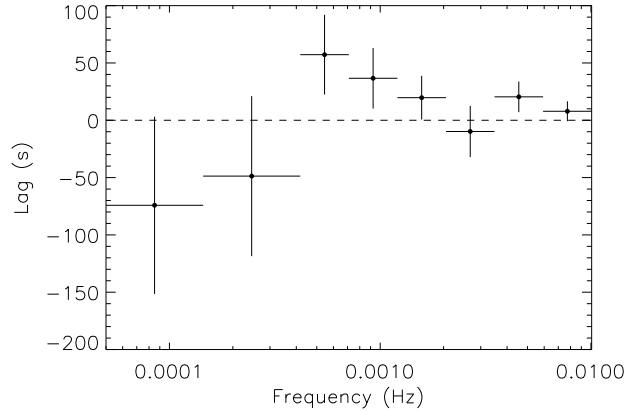}}}
            \caption{The time lag as a function of Fourier frequency between the $0.4-1.5\keV$ and $2-4\keV$ bands.  A negative
lag would imply that the hard band leads the soft.  No significant lag is detected in \1es.
            }
            \label{fig:lag}
            \end{figure}

              Light curves were created in several energy bands between $0.2-12\keV$ to compare the variability at different energies. 
All of the energy bands examined exhibited persistent variability.  
Given the highly variable light curves from \1es\ we decided to perform a search for lags in these data.
To date, over a dozen AGN display reverberation lags 
(e.g., Fabian \et 2009, 2012, Zogbhi \et 2012, Emmanoulopoulos \et 2011, De Marco \et 2011, 2012).
We initially search for lags between two bands, a soft band ($0.4 - 1.5\keV$) where the soft excess is strong, 
and a hard band ($2 - 4\keV$) dominated by the power-law. We extracted light curves in these bands using $20\s$ 
binning. From these light curves we calculate the cross spectrum, and determine the phase lag from the 
argument of the cross spectrum (see Nowak et al. 1999 for a detailed description).  The time lag is simply 
the phase lag divided by $2 \pi f$, where $f$ is the Fourier frequency.
We find no significant lag between these two bands (see Fig~\ref{fig:lag}) even when we combine the first two data points at the lowest frequencies.

\begin{figure}
            \rotatebox{0}
            {\scalebox{0.55}{\includegraphics{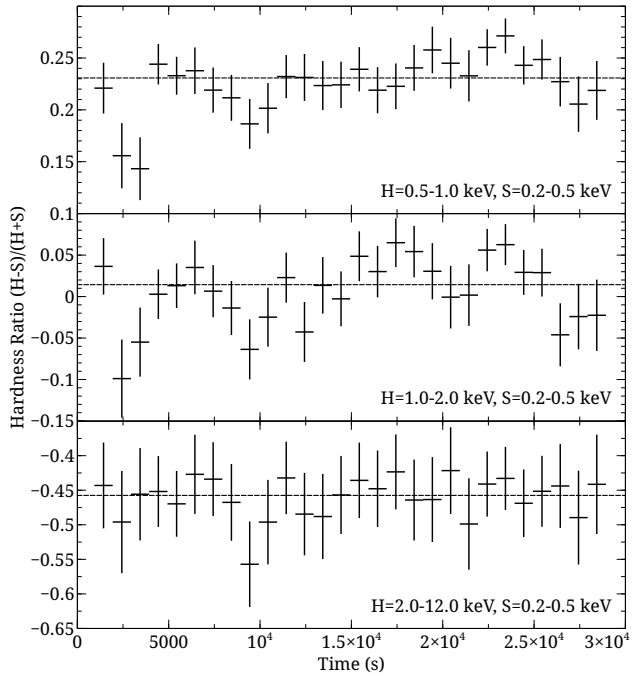}}}
            \caption{The hardness ratio plotted for various energy bands as a function of time.  The light curves are binned in $1\ks$ intervals to improve statistics.  The degree of variability differs depending on the energy bands being compared.
                      }
            \label{fig:hr}
            \end{figure}

            Hardness ratios ($HR = H-S / H+S$; where $H$ and $S$ are the count rates in the hard and soft band, respectively) as a function of time were calculated between all the energy bands.  A few examples are shown in Fig~\ref{fig:hr}.  The degree of variability depends on the energy bands being compared.  The variations were most significant when the intermediate bands were compared to the softest bands (see the upper and middle panels of Fig~\ref{fig:hr}).
                       
\begin{figure}
            \rotatebox{0}
            {\scalebox{0.55}{\includegraphics{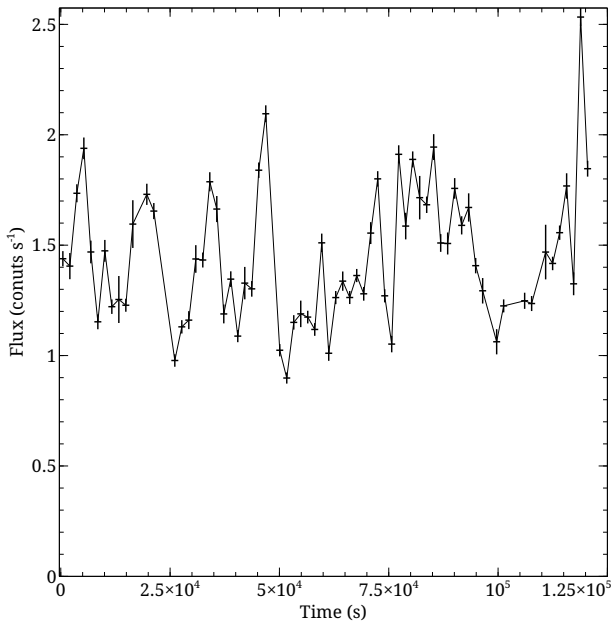}}}
            \caption{The $0.5-10\keV$ \suzaku\ light curve binned in orbital time intervals ($5760\s$).              }
            \label{fig:SuzCurve}
            \end{figure}

            The $0.5-10\keV$ FI CCD light curve is shown in Fig~\ref{fig:SuzCurve} with bins corresponding to orbital 
            time scales ($5760\s$).  The variability is very similar to that exhibited during the \xmm\ observation.  

\begin{figure}
            \rotatebox{0}
            {\scalebox{0.55}{\includegraphics{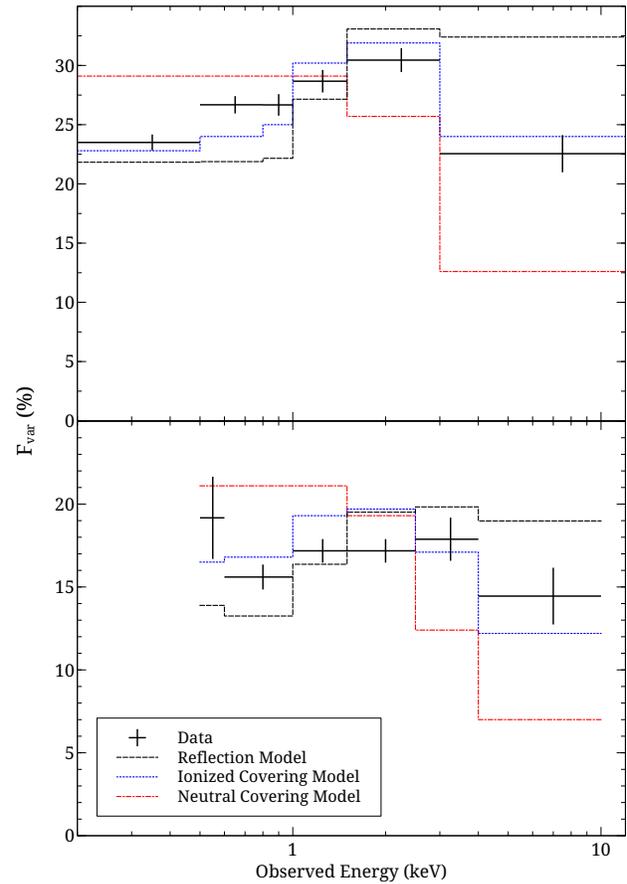}}}
            \caption{Upper panel: The fractional variability spectrum from the \xmm\ observation calculated using $500\s$ binning of the light curves.  The fractional variability expected from the the best-fitting spectral models assuming the fluctuations are dominated by the power law component are overplotted.  Lower panel:  The same as above, but for the \suzaku\ data using $5760\s$ binning of the light curves.          }
            \label{fig:Fvar}
            \end{figure}

            The $\fvar$ in various energy bands is calculated  following Edelson 
            \et (2002) and uncertainties are estimated following Ponti \et (2004).    The \xmm\  light curves are in $500\s$ bins while the \suzaku\ light curves are in $5760\s$ bins.  The $\fvar$ spectrum from the 
            \xmm\ and \suzaku\ observations are shown in Fig~\ref{fig:Fvar}.  
            The spectrum shows the amplitude of the 
            variability peaks at intermediate energies resulting in the bell-shaped $\fvar$ that is commonly seen 
            in unobscured Seyfert galaxies (e.g. Gallo \et 2004; Ponti \et 2010).  

            We considered if the  blurred reflection, neutral partial covering, and ionized partial covering spectral 
            models presented in Section~\ref{sect:physicalModels} could describe the spectral variability seen in 
            Fig~\ref{fig:Fvar}.  In all three cases we considered the possibility that the variability was caused 
            by changing the normalization of the power law component alone.  The models are overplotted on the $\fvar$ 
            spectra in Fig~\ref{fig:Fvar}.  The ionized partial covering describes the variations very well.  
            We note that in unabsorbed sources, the rapid variability is often attributed to variations in the power law component, and it is
            reasonable to expect that such variability is taking place even when line-of-sight absorption is present.  However, variations in covering fraction and/or
            column densities have been observed, even on time scales of less than about $1-2$~days (e.g. Turner \et 2008; Risaliti \et 2005).
            We return to this in Section~\ref{sect:frs} and consider variations in the covering fraction.
            
            Varying the power law normalization in the blurred reflection model gives a reasonable approximation of 
            the $\fvar$ spectrum. The model broadly describes the amplitude,  peak, and general shape of the $\fvar$ 
            spectrum, but there are clearly inaccuracies.

             \subsection {Flux-resolved spectroscopy}
	     \label{sect:frs}	
            
            It stands to reason the variability model (i.e. power law varying in brightness) presented above is an oversimplification.
            For example, in the blurred reflection model, when varying the 
            brightness of the primary emitter (whether by intrinsic fluctuations or via light bending, Miniutti 
            \& Fabian 2004), there should be variations in the dependent parameters of the reflector (e.g. ionisation and reflected flux).  
            
\begin{figure}
            \rotatebox{0}
            {\scalebox{0.55}{\includegraphics{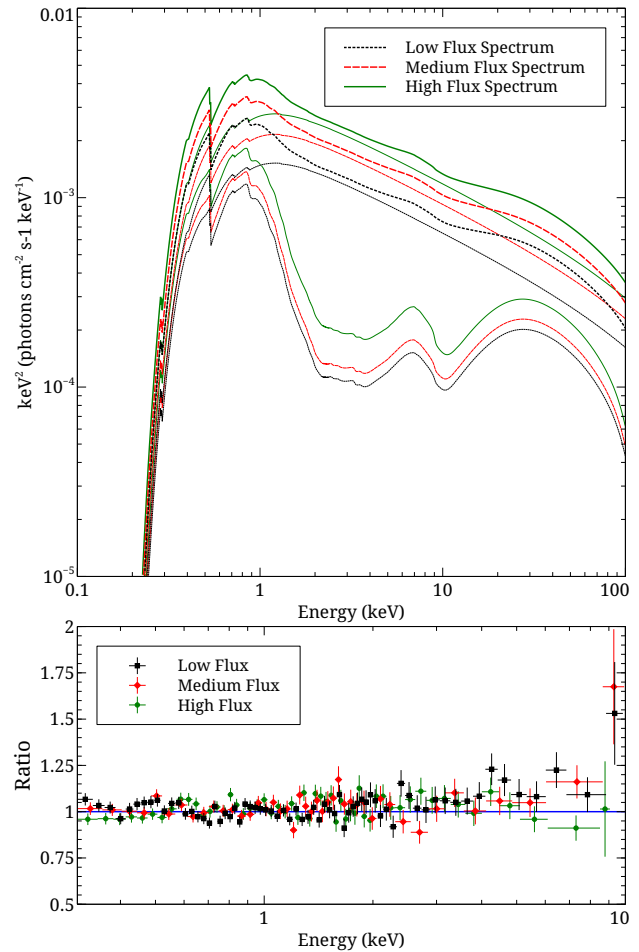}}}
            \caption{Top panel:  The blurred reflection models used to describe the various flux states during the \xmm\ observation.  The differences can be described by changes in the ionisation parameter of the reflector as would arise from changes in the flux of the source illuminating the disc. 
            Lower panel:  The residuals remaining at each flux state based on the model described in the top panel.
            }
            \label{fig:fluxDat}
            \end{figure}

\begin{figure}
            \rotatebox{0}
            {\scalebox{0.55}{\includegraphics{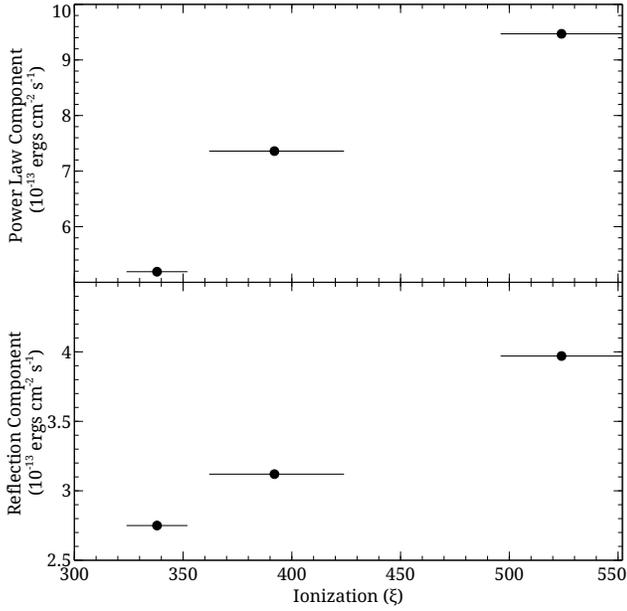}}}
            \caption{The correlation between the ionisation parameter of the reflector and the $20-50\keV$ flux of the power law (top panel) and reflection (bottom panel) components derived from the blurred reflection model description for the \xmm\ flux states (Fig~\ref{fig:fluxDat}).  The correlation follows the expected behaviour that would occur as the flux illuminating the disc varies.
            }
            \label{fig:fluxComps}
            \end{figure}

            To determine 
            if more complicated (but more realistic) variability is required we created flux-resolved spectra of \1es\ in low- ($< 6 \cps$), 
            medium- ($6 - 7 \cps$), and high-flux ($> 7 \cps$) states\footnote{The average, observed  $0.3-10\keV$ flux in each level from low-to-high is approximately $0.8$, $1.1$, and $1.5\times 10^{-11} \ergpscmps$}.  Fitting the three states with the average 
            blurred reflection model, but allowing for changes in the power law normalization resulted in a good fit 
            ($\chidof = 1.04/1421$).  The fit was significantly improved ($\delchi = 83$ for 3 addition free 
            parameters) when we allowed the ionization of the reflector to vary along with the power law. (Fig~\ref{fig:fluxDat})  
            The changes in the ionization are well correlated with the flux in the reflection component (Fig~\ref{fig:fluxComps}) 
            as would be expected (e.g. Miniutti \& Fabian 2004).  The fits to the flux-resolved spectra demonstrate 
            that in terms of the blurred reflection model the variability is more complicated than was assumed for 
            Fig~\ref{fig:Fvar}, and completely consistent with general AGN behaviour.
            
\begin{figure}
            \rotatebox{0}
            {\scalebox{0.55}{\includegraphics{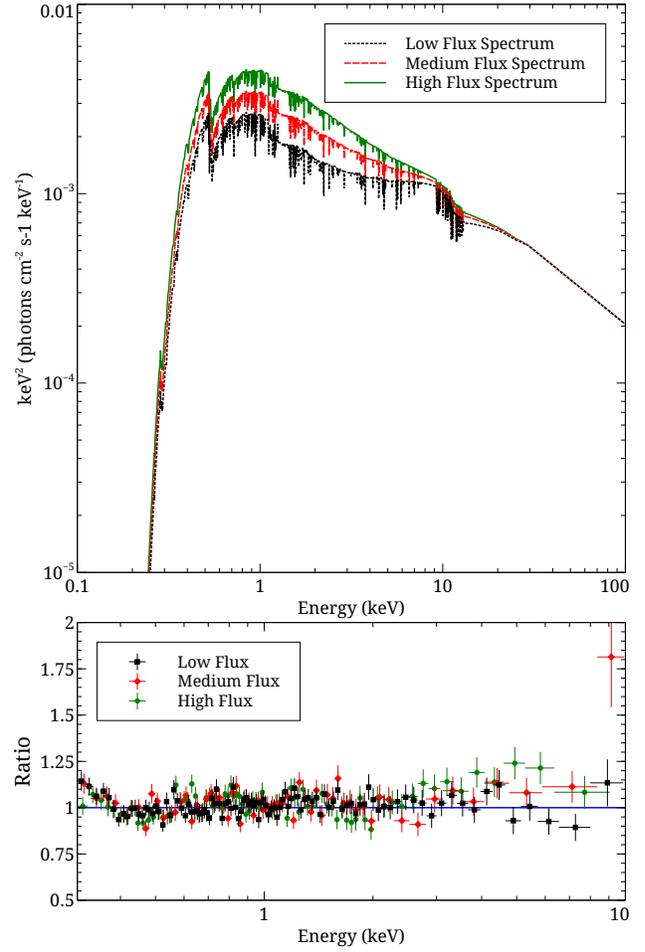}}}
            \caption{Top panel:  The ionised double partial covering model applied to the  various flux states during the \xmm\ observation.  The differences can be attributed to changes in the covering fraction of a relatively constant primary emitter. 
                        Lower panel:  The residuals remaining at each flux state based on the model described in the top panel.
            }
            \label{fig:fluxIPC}
            \end{figure}

	For the ionised partial covering model, we also considered if changes in the covering fraction alone 
	(i.e. constant primary emitter) could describe the various flux states.
	The three spectra could be well fitted ($\chidof = 1.02/1419$; Fig~\ref{fig:fluxIPC}) by allowing only the covering fraction of each absorber to vary.
	In this scenario, the covering fraction of the highly ionised absorber dropped from $\sim 58$ to $\sim 42$ per cent between the high and low state.
	The covering fraction of the colder absorber dropped significantly from $\sim 40$ to $\sim 2$ per cent between the high and low state.



    \section{Discussion}   
    \label{sect:Discussion}

These \xmm\ and \suzaku\ data provide the first high signal-to-noise observations of \1es\ above $2\keV$.  For the first time the existence of a soft-excess below about $1 \keV$ has been confirmed, and an upper limit on a narrow \feka\ line and Compton hump have been established.  
The $0.3-10\keV$ spectra are well fitted with the standard traditional models, in particularly the blackbody plus power law model.  The photon index ($\Gamma \approx 2.3$) and blackbody temperature ($kT \approx 170 \eV$) are both higher than the canonical values, but within observed ranges (e.g. Nandra \& Pounds 1994; Crummy \et 2005).  There is a small level of cold absorption required ($5-10 \times 10^{20} \pscm$) for every model attempted, which could be associated with either the AGN or the host galaxy.

\1es\ exhibits rapid and large amplitude flux and spectral variations.  The bell-shape seen in the $\fvar$ spectrum resembles the typical curve seen in unabsorbed, type-1 Seyferts, in particularly those of narrow-line Seyfert 1 galaxies (e.g. Gallo \et 2004; Ponti \et 2010).   The $\fvar$ spectra and flux-resolved spectra can be well described by either a blurred
reflection model or by an ionized double partial covering model.  In terms of blurred reflection the rapid variability could be described by changing brightness of the primary emitter and corresponding changes in the ionization of the reflector.  Variability based on the ionized partial covering model can be described by changes in the covering fraction while the power law component remains constant.  The variability seems to be in line with what is predicted by each model.

The partial covering models are pure absorption models in that they do not include Compton scattering (e.g. Miller \& Turner 2013) or the fluorescent emission that accompanies photoelectric absorption (e.g. Reynolds \et 2009).   
If the partial covering models could be described in this standard way, then a narrow \feka\ emission line is expected to accompany the absorber.  The denser absorber ($N_H \approx 66 \times10^{22} \pscm$) in the ionized partial covering model would remove about $1.8 \times10^{-5}$ ph~s$^{-1}$ cm$^{-2}$ ionizing photons ($7.08-20 \keV$) from the \xmm\ spectrum.  Assuming the absorber is isotropically distributed around the source, then the strength of the iron line generated from photoelectric absorption is governed by the fluorescent yield.  Adopting the fluorescent yield value for neutral iron (0.347; slightly higher for ionized iron) results in $6.4\times10^{-6}$ ph~s$^{-1}$ cm$^{-2}$ line photons being created.  Such a feature would have an $EW \approx 147\eV$ in the \xmm\ spectrum.  A similar exercise for the \suzaku\ observation predicts a line with $EW \approx 101 \eV$.  Such lines should be detected and are inconsistent with the upper-limit of $30\eV$ found for a narrow \feka\ feature in the data.  We note that Yaqoob \et (2010) argue that Compton-thick lines-of-sight could have much lower \feka\ fluxes (see also Miller \et 2009), rendering our predicted fluxes overestimated.

The expected flux of the Compton bump in the \suzaku\ PIN band can also be estimated from the strength of the predicted iron line using {\tt pexrav}.  The reflection fraction ($R$) is estimated from the equivalent width of the predicted iron line during the \suzaku\ observation: $R = EW / 180 \eV = 0.56$.  The $20-50 \keV$ flux is estimated to be about $1.6 \times 10^{-12} \ergpscmps$, more than double the flux in the pure absorption scenario.  However, the source would still not be bright enough to be detected in the \suzaku\ PIN. 

The ionized double partial covering model provides a good fit to the multi-epoch spectra and the rapid variability.  The variability, both on short and long time scales, can be reasonably attributed to changes in covering fraction and a relatively constant X-ray source.  However, the best-fit is achieved when the absorbers are outflowing at a velocity of $v\approx 0.3c$.   
We note the determination of this blueshift is not driven by any particular spectral feature, but rather by the broadband fit, specifically the soft excess.
The ionised absorption model has been shown to replicate the soft excess in AGN quite well (e.g. Middleton, Done \& Gierli\'nski 2007).  The shape of this
soft excess in a sample of AGN is also well modelled with a black body with a temperature between  $\sim 100-150\eV$ (e.g. Crummy \et 2006; Gierli\'nski \& Done 2004).  The
blackbody temperature measured in  \1es\ was about $170\eV$ (see Table~\ref{tab:traditionalModels}), which is consistent with an average blackbody spectrum seen
in AGN (i.e. $ {\rm kT}\approx130\eV$) blueshifted by $\sim0.3c$.    The outflowing, ionised partial covering seems consistent with the high black body temperature measured in \1es.

If the wind is launched where $v$ is the local escape velocity, then the wind must originate very close to the black hole ($r_{min} \sim (c/v)^2 \rg \approx 10\rg$).   Following Fabian (2012), the kinetic luminosity of the wind, is $L_W = C_f /2  (r_{min}/ \rg) (v/c) N_H / N_T = 0.03 L_{Edd}$, where $N_T = 1.5\times 10^{24} \pscm$ and $C_f$ and $N_H$ are the covering fraction and column density of the denser absorber in Table~\ref{tab:modelData}.   Based on the SED fitting in Section~\ref{subsect:SED}, the Eddington ratio is $L_{bol} / L_{Edd} \ls 0.1$, which would imply that the kinetic luminosity of the wind is
about $\gs 30$ per cent the bolometric luminosity of the AGN.  Such values for $L_W$ do not appear unreasonable for ultrafast outflows according to Tombesi \et (2012).  However, Tombesi \et find the the inner launching radius for such winds are on average $> 10$-times farther from the black hole than $r_{min}$ is in \1es.

The blurred reflection models fit the spectral data quite well.  However we cannot distinguish between extremely relativistic models (e.g. rapidly spinning black hole and compact primary emitter) and less extreme ones.    
The Newtonian model (Blurred reflection (A) in Table~\ref{tab:modelData}) favours about a factor of two overabundance of iron, which is commonly seen in Seyfert X-ray spectra (e.g. Walton \et 2013; Reynolds \et 2012).  The ionization parameter at both epochs is about $350$~erg cm s$^{-1}$ and falls in the range where Auger effect is dominate over fluorescence rendering weak \feka\ features.  
In the relativistic model (Blurred reflection (B) in Table~\ref{tab:modelData}), the iron abundance and ionisation parameters are both lower.  However, the more intense blurring describes the absence of distinct features in the spectrum.  If the soft excess is due to reflection then the inner disc must be present.
The predicted $20-50 \keV$ flux  based on the reflection model is about $1.2 \times 10^{-12} \ergpscmps$ and would not be detectable in the \suzaku\ PIN.   Unfortunately, the null PIN detection of \1es\ alone does not allow us to eliminate any of the potential models.

Both reflection models predict a high inclination ($\sim 85\deg$) consistent with an edge-on disc.  
As with the need for a high velocity outflow in the partial covering models, the preference for a edge-on disc is largely motivated by the ``bluer'' soft excess seen in \1es.  Reflected emission will be more highly beamed in the disc plane thus the reflection component in the spectrum should appear shifted to higher energies.  The high inclination would also be manifested in the blue wing of the \feka\ line, but since the fluorescent line is weak (either due to the Auger effect dominating or extreme blurring), the spectral feature is not significant.

This is intriguing given the Seyfert~2 nature of \1es\ exhibited in the optical band (B03), and arguments that this is a true Sy2 based on optical and near-infrared observations (Panessa \et in prep).  At face-value this would indicate that \1es\ is orientated like a Seyfert~2 in X-rays and optical, but void of the absorption that is associated with the torus.  The small level of cold absorption ($\sim 10^{21} \pscm$) could be from a torus that is currently in some evolutionary state.  This could be confirmed with mid-infrared data to measure the torus emission.

The works of Tran \et (2011) and Wang \et (2012) propose two different scenarios to explain the absence of the BLR in some AGN.  Tran \et predict such objects would be old systems and could be identified by very low $L/L_{Edd}$ values, while Wang \et suggests such objects are young and would exhibit high Eddington ratios.
The SED measurements presented in this work show that \1es\ may be a typical AGN with an Eddington ratio between $0.014-0.11$.  These values appear inconsistent with the predictions of both Tran \et (2011) and Wang \et (2012).  Neglecting the effects of absorption momentarily, if the disc, and consequently the BLR, are seen edge-on as the blurred reflection models suggest, than the Doppler broadening of the BLR lines could be quite extreme.  These  very broad lines could be difficult to distinguish from the optical continuum.  B03 measure the upper-limit on the flux of a broad $H\alpha$ component to be $< 5$ per cent that of the narrow component, but it is not clear what the width of this feature is.  


    \section{Conclusion} 
         \1es\ is a perplexing object that challenges the standard AGN unification model exhibiting optical properties of absorbed systems and X-ray properties of unobscured systems.   According to our SED measurements, \1es\ exhibits a typical Eddington ratio and the absence of its BLR cannot be easily attributed to current models that call for very high or very low $L/L_{Edd}$.  Based on X-ray spectral models we speculate that \1es\ could be an edge-on system (i.e. like a Seyfert 2), with a standard accretion disc that extends to the inner regions, but viewed through a tenuous torus.  Future optical and mid-infrared observations could test this hypothesis.
Future observations with {\it NuSTAR} (Harrison \et 2013) and {\it ASTRO-H} (Takahashi \et 2012) would produce high-quality data above $10\keV$ that would distinguish between the two proposed x-ray models.


    \section*{Acknowledgments}

       The authors are grateful to the referee for valuable comments that improved
        the paper.   LCG would like to thank Kirsten Bonson, Marcin Sawicki and Dave Thompson for discussions.  
         The \xmm\ project is an ESA Science Mission with instruments
        and contributions directly funded by ESA Member States and the
        USA (NASA).  This research has made use of data obtained from the Suzaku satellite, a collaborative mission between the space agencies of Japan (JAXA) and the USA (NASA).



    \bsp
    \label{lastpage}

\end{document}